\documentclass[a4paper]{article}
\usepackage{amsmath}
\usepackage{amssymb}
\usepackage{cite}
\usepackage{fullpage}
\usepackage{numprint}
\usepackage{pifont}
\usepackage{rotating}
\usepackage{xcolor}
\usepackage{hyperref}

\npthousandsep{,}\npthousandthpartsep{}\npdecimalsign{.}

\newcommand{\N}[0]{\mathbb{N}}
\newcommand{\num}[1]{\numprint{#1}}
\newcommand{\yes}[0]{\ding{51}}
\newcommand{\no}[0]{\ding{55}}

\definecolor{yes}{rgb}{0,0.5,0}
\definecolor{no}{rgb}{0.7,0,0}
\definecolor{partial}{cmyk}{0,0,1,0.5}

\renewcommand{\eqref}[1]{\hyperref[eq:#1]{(\ref*{eq:#1})}}
\newcommand{\tabref}[1]{\hyperref[tab:#1]{Table~\ref*{tab:#1}}}

\newcommand{\mysection}[2]{\section{#2}\label{sec:#1}}
\newcommand{\mysubsection}[2]{\subsection{#2}\label{subsec:#1}}
\newcommand{\secref}[1]{\hyperref[sec:#1]{Section~\ref*{sec:#1}}}
\newcommand{\subsecref}[1]{\hyperref[subsec:#1]{Subsection~\ref*{subsec:#1}}}

\newcommand{\includewidetab}[4]
  {\begin{sidewaystable}
     \begin{center}
     \begin{tabular}{#1}
     #4
     \end{tabular}
     \caption{#3}
     \label{tab:#2}
     \end{center}
   \end{sidewaystable}}

\title{Resource Control in P2P Cryptocurrency Networks}

\author{Daniel Kraft \\
XAYA Project \\
Pumpwerkstrasse 5, 8134 Adliswil, Switzerland \\
Email: daniel@xaya.io \\
ORCID: 0000-0002-0862-5350}
\date{August 25th, 2018}

\begin{document}
\maketitle

\begin{abstract}
For decentralised P2P networks, it is very important to have a mechanism
in place that allows the nodes to control resource usage and prevent
flooding and denial-of-service attacks with spam.
In this paper, we discuss and compare the different approaches to fully
decentralised resource control that are used by projects in the
cryptocurrency space.
The introduced methods are then applied to design a decentralised
exchange for Namecoin names (or more generally, crypto assets)
as an example.

\textbf{Keywords:}
P2P Network, Resource Control,
Cryptocurrency, Transaction Fees,
Namecoin, Decentralised Exchange
\end{abstract}

%%%%%%%%%%%%%%%%%%%%%%%%%%%%%%%%%%%%%%%%%%%%%%%%%%%%%%%%%%%%%%%%%%%%%%%%%%%%%%%%
\mysection{introduction}{Introduction}

The launch of the Bitcoin network \cite{bitcoin}
in 2009 fuelled the interest in decentralised P2P networks.
It also enabled such networks for the first time to work with
money and, more generally, \emph{value}.  This made them suitable for an entire
field of new applications that is to date still being explored.
The fundamental property of many of these networks is that they lack any
kind of central point of failure, and enable or even encourage every network
participant to ``verify and not trust''.
But this also means that it is much harder if not impossible by design to
split any processing load of the network.  Instead, typically every node
retrieves and verifies all information,
based on a gossip protocol \cite{gossip}.
Therefore, it is vitally important to restrict the load on the network
to such levels that every node can process and handle
all data; especially so if it is desired that ``ordinary users'' can run
nodes on commodity hardware at home.
(This issue was central to the block-size debate \cite{blocksizeDebate}, that
paralysed the Bitcoin community for quite some time.)

In this paper, we want to review and compare various approaches that can
be used to limit the amount of data sent to and processed by P2P networks
that either directly underpin a cryptocurrency or provide additional services
based on one (like a decentralised cryptocurrency and asset exchange).
While the methods discussed are not new and already employed by existing
projects in the field, we are not aware of any previous work that
collects and compares all of these approaches.

Below, we first discuss the exact goal we want to achieve
with any of the data-limitation methods in \secref{basics}.  We also introduce
basic properties of such methods in that section, which will serve as
the basis for comparisons between approaches.
In \secref{fee-based} and \secref{identity-based}, we introduce the
various methods grouped broadly into ``fee-based'' and ``identity-based''
approaches, respectively.
To conclude this paper, we apply our results to design a
\emph{decentralised exchange for names}
on the Namecoin blockchain \cite{namecoin} as an example
(see \secref{example}).

%%%%%%%%%%%%%%%%%%%%%%%%%%%%%%%%%%%%%%%%%%%%%%%%%%%%%%%%%%%%%%%%%%%%%%%%%%%%%%%%
\mysection{basics}{The Goal and Basic Properties of Resource Control}

For all methods discussed in the following, our goal is always to
\emph{enforce an upper limit to the usage of some network resource}
like node bandwidth, processing power, memory usage or disk space.
In other words, while concepts like a minimum transaction fee
(see also \subsecref{direct fee}) are useful to this end, we will never
explicitly try to enforce a certain transaction fee as the primary goal.
Instead, the fee paid by transactions can be used as a tool to order
transactions by their importance and discard those that do not fit
into the resource limit anymore (through the emergence of a \emph{fee market}).

In Bitcoin itself, the primary limit is the \emph{block size}.  Transaction fees
are used by miners to decide which transactions to put into the scarce space
in a block and which to delay.
The minimum fee itself is dynamic, though, and depends on the
current demand for the existing supply of block space.
Similarly, nodes can decide to relay only transactions that pay a high-enough
fee if they run low on available network bandwidth, and they discard
the lowest-fee transactions from their mempool if its memory usage is too large.

In this context, the actual \emph{transaction fee} serves as a ``knob'' that
can be controlled to achieve the primary goal (reduction in the number of
transactions until they fit into a block).  All data-limiting methods
must have such a knob, but its exact nature can vary.  This is the first
important property that we can use in the following to classify the
different approaches.
In addition, we will use the following two properties for our comparison:

\begin{description}

\item[Blockchain Usage]
As blockchains have inherent scalability limits, block space is a particularly
scarce and expensive resource.  Thus it is important whether the method used
to limit data on some P2P network needs to store data on a blockchain,
and if it does, whether this data usage is the primary goal of the P2P
network anyway (e.g. Bitcoin transactions) or an \emph{additional} cost
that is a consequence of the data-limiting method (e.g. paying a fee in
bitcoin for data sent on a second network, see \subsecref{indirect fee}).

\item[Economic Efficiency]
To ensure that the network resources are allocated to the most important
uses, it should be possible for any user to have their data prioritised
if they are willing to pay enough for it (in some way, not necessarily as a
direct transaction fee).  For instance, if the requirement
is to ``just'' pay a certain fee then this is the case; but if a new user has to
be a member of the P2P network for a month before they are able to send their
data, then the mechanism is not economically efficient.

\end{description}

%%%%%%%%%%%%%%%%%%%%%%%%%%%%%%%%%%%%%%%%%%%%%%%%%%%%%%%%%%%%%%%%%%%%%%%%%%%%%%%%
\mysection{fee-based}{Methods Based on Transaction Fees}

Let us now start to discuss different approaches for limiting data usage
in detail.  In this section, we will look at the first big group of methods:
Those based in some sense on a \emph{fee paid per operation} on the P2P
network.

%%%%%%%%%%%%%%%%%%%%%%%%%%%%%%%%%%%%%%
\mysubsection{direct fee}{Direct Fees}

For networks that have their own cryptocurrency (i.e. cryptocurrency
networks themselves), the most straight-forward way to control the number
of transactions is by imposing a \emph{direct fee}.  In other words,
transactions are required to pay a certain amount $F$ of the network's internal
cryptocurrency in order to be considered for processing.
This payment is typically received by miners to incentivise them for
confirming a transaction, but it could of course also be paid to the
network's operator, a development fund or burnt.

This approach is used in most cryptocurrency networks, including Bitcoin
and Ethereum \cite{ethereum}.
Since the size $s$ of a transaction in bytes is an important
metric approximating the ``overall resource usage'' (in particular, space
used in the ever-growing blockchain), the \emph{fee rate} $f = F / s$
can be used to order transactions by importance.
Miners on the Bitcoin network typically
choose the highest-paying transactions to be included in the available
block space, and nodes can use the fee rate to decide which transactions to
relay and keep in their mempool if they run low on network bandwidth and memory,
respectively.

Of course, the transaction size $s$ is not the only relevant metric that
can be used as basis for the computation of a relative fee.  Ethereum uses
the estimated cost of execution for a transaction, expressed in terms of
\emph{gas}.  For Bitcoin itself, it has been suggested to make the
size of the UTXO set more important rather than using just the
pure transaction size (see, for instance, \cite{feesUtxoSize}).
So the actual quantity used to describe the \emph{cost of a transaction
to the network} can be chosen depending on the particulars.

The \emph{additional} impact that direct fees as control mechanism have
on the blockchain size is limited:  They are applicable only for transactions
on the cryptocurrency network itself, which need to be stored on the blockchain
already.  So the additional cost incurred due to the fee payment is often
negligible, especially if fees are ``implicit'' as is the case for Bitcoin
(where not even an additional transaction output is needed for the fee).

Direct fee payments also ensure that resource control is economically efficient,
since a sufficiently large monetary payment is all that is needed to
prioritise one's transactions; this means that it is easy for everyone to do
so as long as there is enough economic value in the transaction to outweigh
the fee.

%%%%%%%%%%%%%%%%%%%%%%%%%%%%%%%%%%%%%%%%%%
\mysubsection{indirect fee}{Indirect Fees}

For P2P networks that are not a cryptocurrency themselves, one can still
utilise the currency of \emph{another network} to pay fees; we call this
approach \emph{indirect fees}.  Roughly speaking, to send some message $X$
on the primary P2P network, a user has to also send a transaction $T$ on
a secondary cryptocurrency network.  $T$ has to pay a fee in the second
network's native currency; as with a direct fee, this payment can go to
miners, the network operator / developer, or be burnt.
Furthermore, $T$ has to reference $X$ somehow, to prove that the fee being
paid is meant for $X$ and to ensure that a single fee payment cannot be used for
multiple messages on the primary network.
If Bitcoin is used as secondary network, then $T$ could, for instance,
include a hash of $X$ in an \texttt{OP\_RETURN} output; see \cite{dataInsertion}
for a thorough summary of all the available methods to store data in
Bitcoin (those are mostly applicable to other cryptocurrencies as well).

An indirect fee is used, for instance, to control the number of offers
made on the Bisq P2P bitcoin exchange \cite{bisq}.
In order to be allowed to broadcast an offer on the Bisq P2P network, one
has to first create a transaction on the Bitcoin network that pays the
required offer fee.

Using an indirect fee makes it possible to use fee-based resource control
in P2P networks that do not have a native cryptocurrency themselves.
The drawback, however, is that it requires two transactions to be made; this
means that it puts strain on the secondary cryptocurrency network as well as
the primary P2P network and requires blockchain space for the fee transactions
which would not be necessary otherwise.  This means, in particular, that
a user has to pay an \emph{additional} transaction fee on the secondary
network just for paying the fee on the primary one.
For instance, during the usage spike on the Bitcoin network that occurred
in late 2017, the transaction fees necessary to pay the Bisq maker fee were
orders of magnitude larger than the actual Bisq fee itself.

Despite being quite inefficient and costly in that respect, this approach
has a distinct advantage over ``fee-less'' methods (like the ones
discussed later in \secref{identity-based}):  It is possible to use the
fee requirement not just to control resources on the P2P network, but also
to fund development or network operations through the fees.  Bisq uses
the fees to pay for its arbitration system, such that a fee-less method
could not be used.

As with a direct fee payment, an indirect fee is economically efficient---it
allows anyone who sees enough value in a transaction to pay for it, assuming
that the cryptocurrency that is used for fee payment is widely available
and liquid.

%%%%%%%%%%%%%%%%%%%%%%%%%%%%%%%%%
\mysubsection{hashcash}{Hashcash}

Hashcash \cite{hashcash} is an approach that was initially introduced
to combat email spam by requiring senders of emails to perform a proof-of-work
for each message.  It also serves as one important building block for Bitcoin
itself.
True to the original intention of preventing spam, Hashcash can be used to
control resource usage in P2P networks.
This is how it works on a high level:

Suppose Alice wants to send a message $X$ to some P2P network.  She then has
to choose a \emph{difficulty} $d \ge 1$ and find a nonce $n \in \N$
such that
\begin{equation}
\label{eq:PoW}
H(X | d | n) \le \frac{2^b}d.
\end{equation}
Here, $|$ denotes the concatenation operator, $b$ is a number of bits
(e.g. $b = 256$) and $H$ is a hash function with $b$-bit output (e.g. SHA-256).
If we assume that $H$ is not susceptible to preimage attacks, then the
only way to find such $n$ is by brute force.  This means that Alice has to
compute $H$ on average $d$ times to find a valid nonce $n$, and thus
spend potentially a large amount of computation power (and electricity).
Once she has found a suitable nonce, she can broadcast the tuple $(X, d, n)$
to the P2P network.

Bob, who operates a node on the network, now verifies that \eqref{PoW} holds
for each message the node receives.  This can be done very cheaply, since only
one application of $H$ has to be computed per message (independent of $d$).
As resource usage increases for his node, he can set a difficulty target
$\underline{d}$ and only process messages that have $d \ge \underline{d}$---this
gives him a knob that can be used to control the load on his node.

This is the basic method used in Bitcoin's PoW consensus mechanism.  It can,
however, also be used in practice for the original purpose of controlling spam.
See \cite{hashcash} for a list of applications in that spirit; more recently,
it has also been used to control P2P network traffic on the Bitmessage network
\cite{bitmessage}.

One specific property Hashcash for controlling messages in a P2P network
is that it is independent of any cryptocurrency.  A Hashcash PoW can be verified
simply by checking \eqref{PoW}.  Thus it is very simple to
implement, and it does not use any space in a blockchain nor require
transaction fees to be paid for that space.

The main disadvantage, on the other hand, is that computation of a valid
PoW may not be easily doable for ordinary users.  If the threshold
$\underline{d}$ is low enough to make it feasible to solve \eqref{PoW}
on a typical user's computer, then it is likely very cheap to flood the network
with messages using mining rigs with many GPUs or by renting computation
power at some cloud provider.  Conversely, if the threshold $\underline{d}$ is
high enough to prevent this, then PoW's can only be computed by high-end
GPUs or the likes.  Users without such hardware can of course pay someone
else to solve the PoW for them, similar to how Bitcoin miners ``rent out''
their computational power to a mining pool.  But this adds at least another
layer of complexity in the ecosystem, and such a ``market for PoW's'' may
not exist or not be very efficient.
Thus, Hashcash as a method to control resource usage has limited
economic efficiency.

%%%%%%%%%%%%%%%%%%%%%%%%%%%%%%%
\mysubsection{coinage}{Coinage}

In Bitcoin and similar cryptocurrencies, the concept of \emph{destroyed coinage}
can be defined:
For a transaction spending $k$ outputs with values $v_i$ and ages $a_i$,
$i = 1, \ldots, k$, it is
\begin{equation*}
c_{\text{tx}} = \sum_{i=1}^k v_i a_i.
\end{equation*}
In this context, the \emph{age} of an output is the difference between the
timestamps of the spending transaction and the previous transaction
(which created the output).

The maximal theoretical value of $c_{\text{tx}}$ at any time is the
coinage of the full UTXO set,
\begin{equation*}
c_{\text{u}} = \sum_{i=1}^K v_i a_i,
\end{equation*}
where $K$ is now the number of \emph{all unspent outputs}.
This has the interesting property that its time derivative
\begin{equation*}
\dot{c}_{\text{u}}
  = \sum_{i=1}^K v_i \dot{a}_i
  = \sum_{i=1}^K v_i
  \le M
\end{equation*}
is equal to the amount of money currently in the UTXO set and thus
strictly bounded from above by the money supply
(e.g. $M = \num{21000000}$ for Bitcoin).

Because of this, the ``available'' coinage grows at a limited rate.  Thus,
instead of requiring a monetary transaction fee, one can instead require
that each transaction must destroy a certain amount of coinage to limit
the total transaction rate.
This means that coinage can effectively prevent spam attacks while
allowing non-spamming users to send transactions for free (if they are willing
to wait long enough):
If we denote the network's capacity (transactions per second) by $C$,
then a user holding $m$ coins can collect enough coinage to send a transaction
after $M / (m \cdot C)$ seconds on average; the only potential way in which
an attacker can flood the network and prevent this is by using up previously
collected coinage, and thus not sustainable.

This upper bound on the ``cost'' of a transaction is unlike systems based
on a real fee similar to Bitcoin:  There, miners can in theory spend their
earned transaction fees immediately again to send more transactions,
driving up the transaction fee in an almost unlimited way.
(Since each coin can still be spent only once in a single block, there
is a theoretical limit also here.  But it is much higher.)
With coinage, legitimate users can gain an advantage over such an attacker
by waiting longer before spending their coins again (which is the typical
behaviour for non-spamming users anyway).

Bitcoin previously used a coinage-based system to allocate some part of the
available block space and allowed ``priority transactions'' for free.
This was removed for version 0.15, though (in pull request \#\num{9602}),
where it switched to transaction ordering based purely on their fee.
Coinage is also the basis for proof-of-stake consensus methods,
as pioneered by Peercoin \cite{peercoin}.

From the point of view of blockchain-space usage, coinage-based systems for
resource control work similar to fees:  They can be applied quite efficiently
for cryptocurrency networks themselves, where transactions are sent anyway.
But to apply them to non-currency P2P networks, a secondary transaction on a
cryptocurrency network has to be created to ``spend'' the coinage.

Coinage is, however, not very efficient economically:  If someone does not yet
have accumulated coinage to spend, they are forced to wait and collect some
before they can send a message; even if they are willing to pay, they cannot
override this mechanism.  The only potential solution would be a market
for coinage (similar to Hashcash PoW's as described in \subsecref{hashcash}
above); but the feasibility of this depends on the actual method in which
coinage has to be spent, and it is likely hard to build an efficient market.

%%%%%%%%%%%%%%%%%%%%%%%%%%%%%%%%%%%%%%%%%%%%%%%%%%%%%%%%%%%%%%%%%%%%%%%%%%%%%%%%
\mysection{identity-based}{Methods Based on Identities}

The second class of methods that can be used to control resource usage
are not based on ``per-message fees'' but on the concept of \emph{identities}:
Each ``real user'' of the network is granted a certain quota of messages they
may send.  If there is too much load on the network, then this quota is reduced;
and if it is already quite low (e.g. only one message per user per hour),
then the criteria for who is considered a ``real user'' can be made stricter.

While this approach may sound straight-forward and intuitive (matching
perhaps how one would naively design anti-spam measures in a network),
the crux is of course in the definition of ``real users''---particularly
in a decentralised setting.  In the following subsections, we will discuss
several ways to do this; some that won't work in a fully decentralised network
and some that will.

%%%%%%%%%%%%%%%%%%%%%%%%%%%%%%%%%%%%%%%%%%%%%%%%%%%%%%%%
\mysubsection{ip}{Real-Life Identities and IP Addresses}

Perhaps the most naive way to define real users (as opposed to spam bots
and attackers) is through real-life identities, i.e. by ``one real person,
one user''.  But this is, of course, impossible to do in a decentralised
setting.  It would require measures like verification of identity documents
to determine who is a real person, or at least verification of email addresses.
But in a P2P network without any trusted, central instance, this is not
possible to do.  (Furthermore, many decentralised P2P networks aim at least
somewhat to increase user privacy---while an identity verification would
be diametrically opposed to this goal.)

Another, related option is to assign network quota per IP address.  This does
not pose the same privacy problems, and if each node accounts the per-IP quota
it allows by itself, this works without a central instance.  However, it is
still possible for attackers to gain access to many IP addresses and
run Sybil attacks with them.  This problem is also mentioned by Satoshi
in the Bitcoin whitepaper \cite{bitcoin}, where he proposes proof-of-work
to change ``one IP address, one vote'' into ``one CPU, one vote''.
Thus, IP addresses are also not a viable and robust solution to
control resource usage in a fully decentralised setting.

%%%%%%%%%%%%%%%%%%%%%%%%%%%%%%%%
\mysubsection{captcha}{CAPTCHAs}

A common approach to tell ``real users'' and ``bots'' apart in centralised
systems is the use of CAPTCHAs.  Hence it is natural that the idea of using them
in decentralised cryptocurrencies is suggested from time to time,
e.g. in \cite{captchaCoin}.
This, however, can also not work in a fully decentralised network:
There has to be some instance that creates the CAPTCHAs and thus knows the
solution; and this has to be some trusted node, otherwise an attacker can
create the CAPTCHAs to solve themselves or otherwise trick the system.

The project that to our knowledge came closest to using a ``CAPTCHA-like''
system in a cryptocurrency-setting is Motocoin \cite{motocoin}.
This is a cryptocurrency that uses ``proof-of-play'' to reach consensus:
Instead of computing partial hash collisions with hardware, users solve
a puzzle game (set as a motorcycle video game) and submit their winning
sequence of moves as proof.  (This is very similar to how Section~2.4 of
\cite{masteringBitcoin} uses the analogy of solving Sudoku puzzles to explain
how Bitcoin mining works.)
However, while Motocoin was a fascinating, innovative and pioneering
project, it ended up being overrun by bots that could solve the puzzle more
efficiently than humans (mainly by exploiting a loophole in the way
in which the puzzles were created).
Thus, also this approach cannot work in practice;
and even if a puzzle could be found that only humans and not bots can solve,
it would still not be practical to require every user of a P2P network to learn
how to play some puzzle game.

%%%%%%%%%%%%%%%%%%%%%%%%%%%%%%%%%%%%%%%%%%%%%%%%%%%%%%%%%%%
\mysubsection{scarce identities}{Scarce Virtual Identities}

Since real-world identities won't work, we can try for virtual identities
(pseudonyms in a broad sense).  For instance, Bitcoin addresses can be
used as a ``cryptographic identity''.  Of course, that by itself is also
not helpful, as anyone can cheaply generate as many addresses as they wish.
What we really need are virtual identities that are also \emph{scarce
and/or expensive to create}.

One way to create such ``expensive identities'' is through a technique
called \emph{proof-of-burn}:  With Bitcoin and other cryptocurrencies,
it is possible to \emph{provably destroy} coins (e.g. by sending them
to \texttt{OP\_RETURN} or an address for which no private key can exist).
To give weight to a pseudonym or cryptographic key that is used as
their identity, a user can issue one (or
multiple) such burn transactions that reference the pseudonym.
This way, the user has to pay real money to create identities, and can thus
not (cheaply) create many.  Each node in a P2P network can then verify
the burn transactions and grant the corresponding identity a certain quota
of messages it processes.
(A process very similar to this is used by OpenBazaar to associate
reputation to pseudonymous identities, see \cite{openBazaarPoB}.)
The amount of coins that are burnt for any one identity can be used
to further control resource usage, by requiring a larger amount
(thus restricting the set of valid identities) if resources
get scarce.

A related technique is the use of \emph{unspent transaction outputs} (UTXOs)
to create scarcity.  (This is less severe, as it does not require the actual
destruction of coins.)
The basic idea here is that unspent transaction outputs in Bitcoin or
another cryptocurrency are scarce by design, at least if they are required
to have a certain minimum value $m$.  In that case, at most $M / m$ such
outputs can ever exist at any one time,
where $M$ is the maximum available money supply.
Thus, by simply proving that one owns a suitable UTXO, one can claim quota
to send messages in a P2P network.  When that quota is no longer needed,
the output can be spent, so that it does not cost any actual coins
(but still allows the P2P network to control resource limits).
This approach is used, for instance,
to reduce potential spying \cite{jmCommitments} in JoinMarket \cite{joinmarket}.

Both of these related methods require the usage of a secondary cryptocurrency,
but can still be implemented for primary P2P networks that are not currencies
themselves.
To create a proof-of-burn or UTXO, a transaction with associated blockchain
usage and fees has to be sent; however, the resulting
scarce virtual identity can be used
afterwards for an arbitrary amount of time and number of messages on the
primary P2P network, so that no \emph{per-message} fee or blockchain usage
is incurred.  This is a big advantage of this method compared to those
from \secref{fee-based}.
Since all that is required is a one-time payment or just the ownership of coins,
these methods are economically efficient.

%%%%%%%%%%%%%%%%%%%%%%%%%%%%%%%%%%%%%%%%%%%%%%%%%%%%%%%%%%%%%%%%%%%%%%%%%%%%%%%%
\mysection{example}{Example: A Decentralised Exchange for Namecoin Names}

Namecoin \cite{namecoin} is a decentralised name-value storage based on
Bitcoin's protocol (and source code).  It allows its users to register
human-readable names and associate values to them.  Blockchain consensus is
used to establish who was the first to register a name and thus who is
the owner.  While everyone can read values associated to names, only
the owner can change them---this allows for the creation of a decentralised
DNS, among other potential applications.

The owner can also change ownership of a name to an arbitrary Namecoin
address, which either belongs to him as well or someone else.  This allows
to build \emph{exchanges}, where users can trade their names with each other.
In this section, we will discuss a hypothetical fully decentralised
P2P exchange for Namecoin names.
Of course, the focus in the context of this paper
will be on the mechanism to control resource usage and prevent abusive spam.

%%%%%%%%%%%%%%%%%%%%%%%%%%%%%%%%%%%%%%%
\mysubsection{ant}{Atomic Name Trading}

Namecoin transactions, including those that change ownership of a name,
have the same format as transactions in Bitcoin:  The spend one or more
inputs and create one or more outputs.  If a name is changed in the
transaction as well, then one input-output pair is added in addition,
which tracks the name in the same way (with a signature of the previous
owner in the input to authorise the change and the new data in the output).

Due to this structure, one can quite easily trade names for namecoins
\emph{atomically}, i.e. in a single transaction.  This makes it safe to do
in a trustless manner, since either both the name and the corresponding
payment are transferred or none are.

Let's assume that Bob wants to buy a name like \texttt{d/bob} from Sally
for 100~NMC, and that they already negotiated and agreed on the deal.
To finalise the transfer as an \emph{atomic name trade}, the following
steps are performed (see also \cite{ant}):
\begin{enumerate}

\item
Bob prepares a Namecoin transaction that has coins from his wallet as
well as \texttt{d/bob} as inputs.  In the outputs, 100~NMC are sent to
Sally's address, change is sent back to Bob, and the name is updated to list
Bob as owner.

\item
After creating the transaction, Bob signs the inputs that he owns, i.e.
the inputs holding namecoins.  The partially signed transaction is sent to
Sally.

\item
Sally verifies the transaction to make sure it corresponds to the negotiated
deal.  If all is fine, she signs the name input and broadcasts the fully
signed transaction to the Namecoin network.

\item
As soon as the transaction is confirmed, the trade has been completed.
If the transaction does not get confirmed for any reason (e.g. one of the
parties performs a double spend), then \emph{none} of the transfers happen.

\end{enumerate}

%%%%%%%%%%%%%%%%%%%%%%%%%%%%%%%%%%%%%%%%%%%%%%%%%%%%%%%%%%%%%%%%%%
\mysubsection{example design}{Brokering Trades with a P2P Network}

So far, atomic name trading as described in \subsecref{ant} is only one
part of a name exchange:  We also need a way to broker trades between
parties that are not yet in contact (or even aware of each other).
For this, we can use a gossip P2P network to exchange offers.
It could work like this:
\begin{enumerate}

\item
Sally and everyone else who want to sell a name send a message through
the network, announcing their offer:  Which name they want to sell,
what the asking price is and their Namecoin address to receive the payment,
at a minimum.
To prevent false offers, she has to sign her offer with the private key
corresponding to the Namecoin address that currently owns the name.

\item
Each node in the exchange network keeps track of the currently active offers
and can show them to the user through some UI.

\item
Bob and others can browse through the offers on their node.
When he finds an offer that he wants to take, he prepares the partially
signed transaction as described above
and sends it again through the network.

\item
Sally receives Bob's offer and potentially other offers as well
on her P2P node.  She can now pick the one she prefers (if any),
sign her part and broadcast it to the Namecoin network.

\end{enumerate}

This design is quite simple; however, one of the main issues that still need
to be addressed is \emph{how to prevent flooding the network with spam offers}.
Let us now discuss which of the methods for resource control that were
introduced before can be used in this situation.

First, let us consider fee-based methods as in \secref{fee-based}:
A direct fee is not applicable, since the exchange network is not
a cryptocurrency itself.  An indirect fee could be used similar to
how it is done in Bisq, but that is wasteful and costly as it requires
each offer to be accompanied by a Namecoin transaction.
Similarly, coinage-based resource limits would also require separate Namecoin
transactions and are thus also not a good idea.
Hashcash could be used similar to Bitmessage, but that has its own
drawbacks (see \subsecref{hashcash}).
So none of the fee-based methods are really suitable here.

We can, however, very well use a method based on scarce virtual identities
(see \subsecref{scarce identities}).  In fact, both sellers and buyers
have already a ``scarce identity'' based on an UTXO:  For the seller,
the name's current output can act as such; for the buyer, the transaction
inputs paying the name's price.
So a very simple way to limit the rate of offers in the network is to
give each name and each Namecoin UTXO of a certain minimum size (e.g. 1~NMC)
permission to send one offer (or partially signed transaction) to the network
per hour (or day).  Each of the network's nodes tracks the quota
for itself, and prohibits messages that go over it.
A rate limit like this will not impact legitimate users of the exchange
at all, but immediately prevent spamming as UTXOs are limited.
If that is not enough, then nodes can increase the minimum size of UTXOs
they require to further limit the rate of messages they process.

%%%%%%%%%%%%%%%%%%%%%%%%%%%%%%%%%%%%%%%%%%%%%%%%%%%%%%%%%%%%%%%%%%%%%%%%%%%%%%%%
\mysection{conclusion}{Conclusion}

In this paper, we have introduced different methods that are used in
existing cryptocurrency (and cryptocurrency-related) P2P networks to
limit resource usage and prevent flooding attacks.  These methods have
different properties, so that they are suitable in different situations.
\tabref{summary} is a summary of our comparison.

For cryptocurrencies themselves, direct fees (as used by the major networks)
are a good and straight-forward method to limit the number of transactions
that are processed.
For networks that are not a cryptocurrency it gets more complicated:
In that case, usage of an explicit fee introduces the need to pay the
fee on another network in a separate transaction, which adds \emph{extra}
blockchain usage and costs.  For those networks, methods based on
scarce virtual identities as discussed in \subsecref{scarce identities}
can make a lot of sense.

\includewidetab{ll|c|cc|cc}{summary}
  {Comparison of the different methods introduced in \secref{fee-based}
   and \secref{identity-based}.}
  {
    \textbf{Method} & \textbf{Class}
      & \textbf{Controlling Quantity}
      & \textbf{Currency} & \textbf{Non-Currency}
      & \textbf{Blockchain Usage} & \textbf{Economic Efficiency}
      \\ \hline \hline
    \textbf{Direct Fee} & Fee-Based & Relative Fee
      & \textcolor{yes}\yes
      & \textcolor{no}\no
      & \textcolor{partial}{Yes}
      & \textcolor{yes}{Yes} \\
    \textbf{Indirect Fee} & Fee-Based & Relative Fee
      & \textcolor{no}\no
      & \textcolor{yes}\yes
      & \textcolor{no}{Additional}
      & \textcolor{yes}{Yes} \\ \hline
    \textbf{Hashcash} & Fee-Based & PoW Difficulty
      & \textcolor{yes}\yes
      & \textcolor{yes}\yes
      & \textcolor{yes}{No}
      & \textcolor{partial}{Limited} \\ \hline
    \textbf{Coinage (direct)} & Fee-Based & Relative Destroyed Coinage
      & \textcolor{yes}\yes
      & \textcolor{no}\no
      & \textcolor{partial}{Yes}
      & \textcolor{no}{No} \\
    \textbf{Coinage (indirect)} & Fee-Based & Relative Destroyed Coinage
      & \textcolor{no}\no
      & \textcolor{yes}\yes
      & \textcolor{no}{Additional}
      & \textcolor{no}{No} \\ \hline \hline
    \textbf{Proof-of-Burn} & Identity-Based & Quota, Burnt Amount
      & \textcolor{yes}\yes
      & \textcolor{yes}\yes
      & \textcolor{yes}{Only Setup}
      & \textcolor{yes}{Yes} \\
    \textbf{UTXO} & Identity Based & Quota, UTXO Amount
      & \textcolor{yes}\yes
      & \textcolor{yes}\yes
      & \textcolor{yes}{Only Setup}
      & \textcolor{yes}{Yes} \\
  }

To showcase the methods on an example, we designed a P2P network that can
be used as a \emph{decentralised exchange} for Namecoin names.  In that
situation, we found that the usage of scarce identities based on UTXOs
makes the most sense, as UTXOs are already directly involved in the
name trading process.
While this example is mainly intended to compare the different methods,
we believe that such a decentralised name
exchange could also be useful to build in practice; furthermore, the design can
be easily adapted to build, for instance, a more general decentralised exchange
for cryptocurrency assets.

%%%%%%%%%%%%%%%%%%%%%%%%%%%%%%%%%%%%%%%%%%%%%%%%%%%%%%%%%%%%%%%%%%%%%%%%%%%%%%%%

\bibliography{references}{}
\bibliographystyle{plain}

\end{document}